Temporal evolution of adherents of the major religions in Mexico: Avrami model application-Kolgomorov solid training.

Mauricio González Aviles, Servin HermelindaCampuzano

Universidad Intercultural Indígena de Michoacán (UIIM)

Mauricio Gonzalez Aviles

gamauricio@gmail.com

Servin Hermelinda Campuzano

merlysc@gmail.com

**Phone:** (434) 342-23-07, 342-55-32

**Summary.** It applies a mathematical model of solid formation, the model of Avrami-Kolgomorov [Ausloos & Petroni, 2007] to model the time evolution of percentage of adherents of the major religions practiced in Mexico, adjusting the corresponding parameters with available records in the period from 1950 to 2000 [Molina-Hernandez, 2003; INEGI, 2005]. A comparison is made with the application of the model to global trends and concludes that Catholicism is in a marked disaggregation and trends of Christianity in Mexico are similar to global.

1.    *Introduction*

The purpose of this paper, from the perspective of "sociophysical", is to explain some aspects regarding the historical evolution of percentage of adherents of the major religions in Mexico. The sociophysical is a new interdisciplinary branch of physics that advocates the use of methods and concepts of physics used in complex systems, for the study of collective interactions in societies and not intended to be a mere application of quantitative methods or mathematicians, but a new conception of social phenomena from an interdisciplinary perspective, as "emergent" properties of a set of individuals who interact with each other to produce new behaviors that can not be reduced to the study of the individual components [Galam, 2004]. Since it is a discipline and a new point of view, is still in its infancy, so focusing, right now, in the search for general patterns of social behavior. As you progress the development of the theory and the interaction between physicists, mathematicians and sociologists, is expected to develop and implement appropriate experiments to processes, and provides ways to contrast the ideas, models and theories created. Sociophysical Specifically, main objective, mathematical modeling of large-scale social phenomena using statistical physics.

To identify and clarify the issues under study, it is necessary to consider certain general considerations as starting premises, around this aspect of the society. Therefore, it is established that religions develop in their beliefs and are adapting to changes in society, in this sense, religion can be studied as a variable in social organization, in the same way, such as , competition between languages or spread of a disease. recently the dynamics of languages used in the world has been studied with some interest, especially in the case of extinction processes them due to competition with other languages [Abrams & Strogatz, 2003]. Therefore, in principle it is plausible to consider whether these considerations apply to the case of religions. It should be explained that aspects such as the origin of religion, history or finding hierarchies, not under consideration from this point of view. Nor will enter discussions on the definition of a religion, it is recognized that there are several names that, in fact, may hinder the collection of data and its subsequent analysis. However, a similar problem exists in addressing the case of languages. Indeed, in principle it might be thought that there are many similarities in the mathematical modeling sense, however, as discussed below, significant differences are found, for example, the fact that the models can be considered to be bilingual but not profess two religions at once.'simportant to emphasize the fact that it is considered as a parameter and physical object of study, the number of adherents of religions and no religions themselves. To address the problem, we have followed the model used in [Ausloos & Petroni, 2007], in which religious adherence has been considered as degree of freedom, in the sense of statistical physics, for a "human agent" that belongs to a certain population. The evolution of the distribution and duration of religions can thus be studied as a set of agents that act reciprocally and heterogeneously.'sPossible to consider, from a point of view "macroscopic": How many religions exist a given time?, Or a point of view "microscopic": How many adherents have a religion in a given time? Does it increase or decrease the number of adherents?, And how?.

Differential equation Avrami-Kolgomorov, which usually describes solid state transformations, such as crystal growth, is used to set parameters for the model, this is discussed in greater detail in section (3). In [Ausloos & Petroni, 2007] applies the model to the case of the "major" religions in the world.

The aim of this work is the application of the model [Ausloos & Petroni, 2007] to the case of the Mexican population in terms the percentage of Catholics, Protestants and no religion, with respect to the total population. For purposes of comparing trends has been considered as adherents to Christianity to the sum of Catholics and Protestants in a given time.

The main specific objectives of this study are:

To apply the model of evolution in time, the case of Mexico, to analyze trends according to its composition of religious adherence, that is, as Catholics, Protestants, without religion and Christianity (Catholics over Protestants). In the same way, to make projections for the future. Compare trends in percentages of adherents of different religions with global trends studied in [Ausloos & Petroni, 2007].

The paper is organized as follows:

1. Description mathematical model and the plausibility of its use.
2. Immediately explains its application to the case of adherents of religions in Mexico and the results of such application.
3. Dbtained results are discussed and future projections closest pair.
4. Conclusions.

2. *Mathematical model of evolution*

Temporalmodel considered in [Ausloos & Petroni, 2007] proposed to simulate the evolution of the number of adherents of religions, a process of nucleation-growth-death, in analogy with crystal growth studies [Cloots et al., 1996]. The nucleation can be understood as the beginning of a change of state in a small but stable region. Avrami equation [Avrami, 1940] describes how solids are transformed from one phase to another. May specifically describe the kinetics of crystallization, it can be applied generally to other phase change materials, such as rates of chemical reactions, and can even be used in analysis of ecological systems. According to the study done in [Ausloos & Petroni, 2007] a differential equation can be used to explain the evolution of the number of adherents (in percentage terms relative to the world population) of the major religions of the world, in terms of competition among entities within the meaning of the Avrami equation.

$$\frac{dg(t)}{dt} = \gamma t^{-h}[1 - g(t)]. \qquad (1)$$

Where $t$ is the variable that represents time, $g(t)$ represents the percentage of adherents of any religion with respect to the world population, $\gamma$ is a parameter to be determined (is positive for If growth processes) and $h$ represents a measure of the growth process (or decrease) in the continuous-time approximation.

This equation is solved trivial (linear differential equation of first order), obtaining the evolution of percentage of adherents as continuous function of time

$$g(t) = 1 - \eta e^{\frac{-\gamma}{1-h} t^{1-h}} \qquad (2)$$

which $\eta$ is related to the initial condition.

As the $h$ parameter, according to the standard model of crystal growth [Avrami, 1940] must be positive and less than 1, that is, if defi $n = 1 - h$ ned, then $n$ itshould be less than otherwise indicates the possibility of the separation of crystal structure. For the application in terms of adherents to religions, $n$ if less than 1, would imply an increase in the percentage of these, otherwise, the breakdown of adherents of any religion. In the case where the absolute value $h$ is not within the $(0,1)$ range, it can be considered that the growth process is heterogeneous nucleation and / or to guess which is due to influences from any external field. Moreover it should be noted that when $h$ is greater than 1, the solution decays Avrami equation with respect to a maximum at a certain time, this interpreted in terms of religions give information when a religion has reached the maximum percentage adherents, but in reality, the information available is very difficult to establish precisely when such a situation has occurred.

## 3. *Application of the model to the case of adherents of religions in Mexico*

Data religones the professors of the Catholic and Protestant, and as those who profess no religion were taken from [Molina-Hernandez, 2003] and [INEGI, 2005].

In [Molina-Hernandez, 2003, p. 93-94] is meant by religion "as a set of social practices guided by beliefs supramundane forces to which the believer recognizes their power to regulate the actions of humans, usually including a specific religious institution, and susceptible classification, quantification and spatial location through religious affiliation declared by the individual and registered by the population census. "Taking the percentage of professors of religion in Mexico with respect to the total population, the model has been applied [Ausloos & Petroni, 2007], in which we have obtained the best settings depending on the parameters involved in the model for this function is minimized (least squares approach), $E = \Sigma_i |g(t_i) - gd_i|^2$ where $g(t)$ is the solution of equation ( 1 ) a $gd_i$ nd the data are taken from [Molina-Hernandez, 2003; INEGI, 2005]. To make the adjustment has been considered an interval and a partition of 1000 points for each of the three parameters of the model.

The best settings for the parameters are summarized in Table (1):

| RELIGION | $\eta$ | $\gamma$ | $h$ |
|---|---|---|---|
| CATHOLICS | $2.0400 \times 10^{-2}$ | $-9.9999 \times 10^{-3}$ | $-0.2845$ |
| PROTESTANT | $0.9930$ | $3.9999 \times 10^{-4}$ | $-0.3787$ |
| WITHOUT RELIGIÓN | $0.9969$ | $1.0000 \times 10^{-3}$ | $0.1270$ |
| CHRISTIANITY | $0.1310$ | $-3.4660$ | $1.5020$ |

*Box 1. It shows the fit of the model parameters.*

In the following figures, the graphs also show the available data and model approach curves. In all four graphs the horizontal axis represents the time in years and the vertical axis the percentage of adherents. The available data are represented by black dots and thick continuous lines represent the model resulting function. We have plotted the four cases under consideration: Catholics, Protestants, without religion and Christianity (Catholics over Protestants).

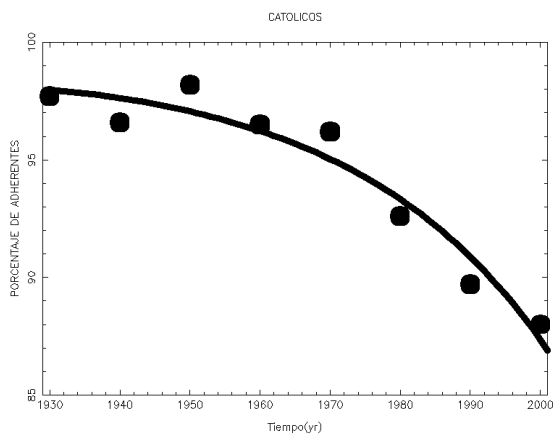
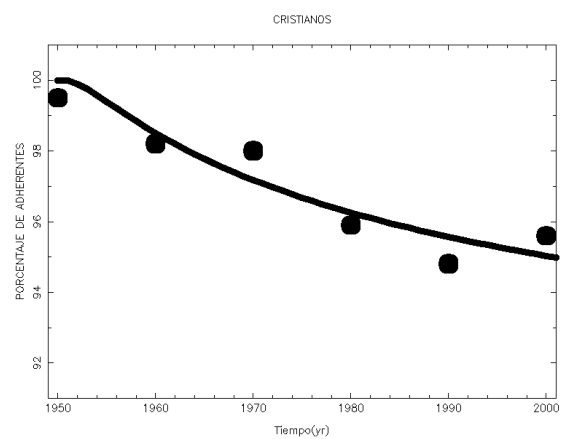

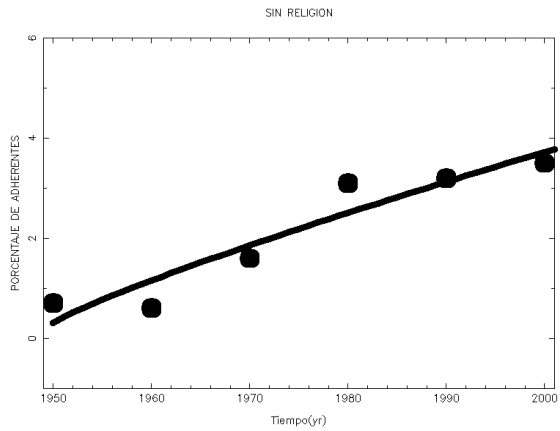 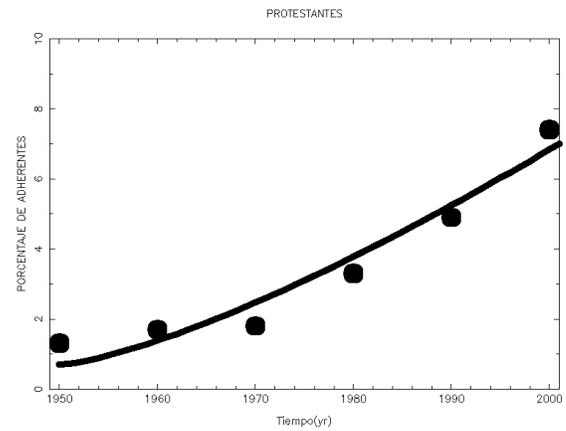

1.

## *Discussion of Results*

According to the results summarized in Table 1 shows that for the case of Protestants without religion the parameter value $\gamma$ is positive, therefore, the percentage of adhesive as a function of time and is growing can be observed also in the corresponding graphs. Otherwise, for the percentage of Catholics and Christianity (Catholics over Protestants), the value of this parameter is negative and implies a decrease in the percentage of adherents over time, as also confirmed in the corresponding graphs.

As to the values of th $n=1-h$ e parameter, for the case of Christianity in which the absolute value is in the interval $(0,1)$ and interpreted that evolution is heterogeneous and / or is conjectured to have a strong dependence on interactions or external pressures. For those without religion is presented a case of standard growth according to the model, since iti $h$ s positive and less than. In the case of the Catholic adherents gives a parameter $h$ which is interpreted as the possibility of separation of the crystalline structure, which in this context implies that the evolution is in a disaggregation process.

The values calculated for $h=1.5020$ the case of Christianity (Catholics over Protestants) are comparable to the global case [Ausloos & Petroni, 2007]. And besides this value $h$ is greater than 1, which translates to the maximum adhesive has been found to have in the past.

## 4. *Conclusions*

According to the application of a mathematical model of solids formation, the number of adhesive applied major religion in Mexico, found that:

1. The percentage over time of adherents of Protestantism without religion is growing. The case of no religion is presented as a case of standard growth according to the values found for the parameters of the model.
2. Decrease is in the case of Catholicism and Christianity (Catholics over Protestants).
3. According to the setting parameters for Catholics, we find that the percentage of members decreases and is in a state of disintegration.
4. For Christianity found its adherents maximum percent is in the past. One can surmise that its decrease is heterogeneous and / or rely on interactions and / or external pressures.
5. Finally theparameter value $h$ associatedis comparable in the case of the global trend, which is considered to have some degree of stability through time.

## 5. *Acknowledgments*

We thank the facilities given to the preparation of this work to the Universidad Intercultural Indígena de Michoacán (UIIM).